\documentclass[10pt,twocolumn,brazil,english,amsmath,amssymb, aps]{revtex4}
\pdfoutput=0 
\usepackage[T1]{fontenc}
\usepackage[latin9]{inputenc}
\usepackage{amssymb}
\usepackage{graphicx}
\usepackage{esint}

\makeatletter

\@ifundefined{textcolor}{}
{%
 \definecolor{BLACK}{gray}{0}
 \definecolor{WHITE}{gray}{1}
 \definecolor{RED}{rgb}{1,0,0}
 \definecolor{GREEN}{rgb}{0,1,0}
 \definecolor{BLUE}{rgb}{0,0,1}
 \definecolor{CYAN}{cmyk}{1,0,0,0}
 \definecolor{MAGENTA}{cmyk}{0,1,0,0}
 \definecolor{YELLOW}{cmyk}{0,0,1,0}
}

\usepackage{dcolumn}

\makeatother

\usepackage{babel}
\begin{document}

\title{The impact of dark energy perturbations on the growth index}

\author{Ronaldo C. Batista}

\email{rbatista@ect.ufrn.br}

\selectlanguage{english}%

\affiliation{Escola de Ci\^encias e Tecnologia, Universidade Federal do Rio Grande
do Norte, Caixa Postal 1524, CEP 59072-970, Natal, Rio Grande do Norte,
Brazil.}

\date{\today}
\begin{abstract}
We show that in clustering dark energy models the growth index of linear
matter perturbations, $\gamma$,	can be much lower than in $\Lambda$CDM or
smooth quintessence models and presents a strong variation with redshift.
We find that the impact of dark energy perturbations on $\gamma$ is enhanced
if the dark energy equation of state has a large and rapid decay at low redshift.
We study four different models with these features and show that we
may have $0.33<\gamma\left(z\right)<0.48$ at $0<z<3$. We
also show that the constant $\gamma$ parametrization for the growth
rate, $f=d\ln\delta_{m}/d\ln a=\Omega_{m}^{\gamma}$, is a few percent
inaccurate for such models and that a redshift dependent parametrization
for $\gamma$ can provide about four times more accurate fits for
$f$. We discuss the robustness of the growth index to distinguish
between General Relativity with clustering dark energy and modified
gravity models, finding that some $f\left(R\right)$ and clustering
dark energy models can present similar values for $\gamma$.
\end{abstract}
\maketitle

\section{Introduction}

The understanding of the accelerated expansion of the universe is one
of the greatest challenges in physics. It may be caused by a yet unknown
form of energy with negative pressure, generically called dark energy,
or by a new theory of gravity and space-time. Since different theoretical
scenarios and their various models are designed to reproduce the expansion
history, from the observational point of view we certainly need to
analyze the evolution of cosmological perturbations in order to be able
to falsify the models and have a better indication of what drives
the accelerated expansion.

One particular simple and powerful tool to study the linear growth
of cosmic structures and discriminate between the various theoretical
possibilities is the growth rate of matter perturbations, $f=d\ln\delta_{m}/d\ln a$,
which is usually parametrized by $f=\Omega_{m}^{\gamma}$, where $\Omega_{m}$
is the matter density parameter and $\gamma$ is the so-called growth
index. It is well-known that $\Lambda$CDM and quintessence models
(minimally coupled canonical scalar fields) can be well described
by a scale independent and constant $\gamma\simeq0.55$ \cite{Wang:1998gt,Linder:2005in,Linder2007}
with very small corrections \cite{Tsujikawa2013}. Since the underlying
theory of gravity in such models is the General Relativity, it is
usual to refer to $\gamma_{GR}\simeq0.55$ as the value of the growth
index in General Relativity. On the other hand, modified gravity models
such as $f\left(R\right)$ models may have $0.40\lesssim\gamma\lesssim0.43$
at $z=0$, besides being scale dependent \cite{Tsujikawa2009,Gannouji2009},
and Dvali-Gadabadze-Porrati (DGP) braneworld gravity model \cite{Dvali2000b}
has $\gamma\simeq0.68$ \cite{Linder2007}. It has been shown by reference
\cite{Heavens2007} that, ongoing observational projects, like Dark
Energy Survey \cite{Abbott:2005bi}, should be able to distinguish a
deviation of $0.179$ from $\gamma_{GR}$ and future space based projects,
like \textit{Euclid} \cite{Laureijs2011}, a deviation of $0.048$,
which proves the power of the growth index to falsify the various
models of cosmic acceleration. 

In the context of quintessence models, dark energy perturbations are
relevant for dark matter growth only on Hubble scales and then are
usually neglected on small scales, which are the observationally relevant
for the determination of $f$. This is due to the fact that quintessence
perturbations have unitary effective sound speed (in its rest frame),
$c_{{\rm eff}}^{2}=\left(\delta p/\delta\rho\right)_{{\rm rest}}=1$
\cite{Hu1998c}. Hence its sound horizon, $\sim c_{{\rm eff}}H^{-1}$,
is of the order of the Hubble radius and dark energy perturbations
are strongly suppressed on smaller scales. However, there are models
of dark energy based on non-canonical scalar fields in which the effective
sound speed is variable and even negligible \cite{Chiba2000,Armendariz-Picon2001,Chimento2005c,Creminelli2009,Lim2010}.
When $c_{{\rm eff}}\ll1$ dark energy perturbations grow at the same
pace as matter perturbations \cite{Abramo:2008ip,Sapone2009} and
can be as large as them when $w=p_e/\rho_e\simeq0$ \cite{Batista:2013oca},
like during the matter dominated period in Early Dark Energy models.
Hence it is possible that in such models the growth of matter perturbations
is quite different from quintessence or $\Lambda$CDM models. In this
case, we have to ask whether the values of $\gamma$ could be mistaken
for modified gravity models.

The issue of how to differentiate a new energy component from a new
gravity theory has been discussed a lot in the literature, e.g., \cite{Kunz2007,Jain2008,Bertschinger2008,Dossett:2013npa}.
Although most smooth dark energy models can be distinguished from
modified gravity models, if dark energy can cluster this task can
be much more difficult. However, reference \cite{Dossett:2013npa}
has recently claimed that clustering dark energy models do not impact
the growth index severely, and its value remains near the $\Lambda$CDM
one, so these models could be easily distinguished from modified gravity.
Moreover, it was found that clustering dark energy models with constant
equation of state differ only about $5\%$ from $\gamma_{GR}$ \cite{Ballesteros2008}. 

In this paper we will show that if clustering dark energy is allowed
to have a large and rapid decay of its equation of state at low redshift,
a situation that was not considered in references \cite{Ballesteros2008,Dossett:2013npa},
then dark energy perturbations will strongly impact the growth index
and it would be difficult to distinguish this model from some $f\left(R\right)$
models.

Although these features may seem unnatural in the context of a single canonical scalar field model, 
there are at least two other possible ways to construct models with this evolution. One can construct a k-essence model requiring negligible effective sound speed \cite{Creminelli2009} and tracking behavior \cite{Chiba2002}, which can in principle induce a fast transition of the equation of state at low redshift. It is also 
possible to make use of two scalar fields, one of which is a Lagrange multiplier that enforces
the sound speed to be exactly zero, regardless of the background evolution \cite{Lim2010}, which can also present the tracking behavior. For the purpose of this paper, however, it will be sufficient to describe dark energy as a perfect fluid with a parametrized evolving equation of state and a constant effective sound speed. 

This paper is organized as follows. In section II we present the background
evolution of four dark energy models that we will use to illustrate
the impact of dark energy perturbations on the growth index. In section
III we present the equations which we use to evolve the linear perturbations.
The results for the growth index and the comparison with some modified
gravity models are shown in section IV. We present the conclusions
in section V.

\section{Background evolution}

We study a background evolution given by a flat FRW model with pressureless
matter (dark matter plus baryons) and dark energy with equation of
state parameter $w=p_{e}/\rho_{e}$, then Friedman equations are given
by: 

\begin{equation}
\mathcal{H}^{2}=\frac{8\pi G}{3}a^{2}\left(\rho_{m}+\rho_{e}\right)
\end{equation}
and
\begin{equation}
\dot{\mathcal{H}}=-\frac{4\pi G}{3}a^{2}\left[\rho_{m}+\rho_{e}\left(1+3w\right)\right]\,,
\end{equation}
where $\mathcal{H}=\dot{a}/a$, the dots represent the derivative
with respect to conformal time. As usual, matter and dark energy density
parameters are given by $\Omega_{m}\left(a\right)=\rho_{m}\left(a\right)/\rho_{c}\left(a\right)$
and $\Omega_{e}\left(a\right)=\rho_{e}\left(a\right)/\rho_{c}\left(a\right)$,
where $\rho_{c}$ is the critical density, $\rho_{c}\left(a\right)=3H^{2}/8\pi G$,
and $H=\mathcal{H}/a$ is the Hubble parameter as a function of the
physical time. 

Reference \cite{Dossett:2013npa} analyzed dark energy models using
the Chevallier-Polarski-Linder (CPL) parametrization \cite{Chevallier:2000qy,Linder:2002et}: 

\begin{equation}
w=w_{0}+w_{a}\left(1-a\right)\,.
\end{equation}
Although it can accurately describe the background evolution of many
quintessence models and some modified gravity models, it can not reproduce
a rapid transition of $w$ at low redshift. Incidentally, as we will
show, if $w$ is allowed to have a rapid transition at low redshift,
dark energy perturbations may strongly impact the growth of matter
perturbations. In order to demonstrate this we will use the following
parametrization \cite{Corasaniti2003}: 
\begin{equation}
w\left(a\right)=w_{0}+\left(w_{m}-w_{0}\right)\frac{1+\mbox{exp}\left(\frac{a_{c}}{\Delta_{m}}\right)}{1+\mbox{exp}\left(-\frac{a-a_{c}}{\Delta_{m}}\right)}\frac{1-\mbox{exp}\left(\frac{a-1}{\Delta_{m}}\right)}{1-\mbox{exp}\left(\frac{1}{\Delta_{m}}\right)}\,,\label{eq:corass_parm-1}
\end{equation}
where $w_{m}$ is the limiting value of $w$ in the matter dominated
era, $w_{0}$ is the value of $w$ now ($a=1$ or $z=0$), $a_{c}$
is the moment of the transition from $w_{0}$ and $w_{m}$ and $\Delta_{m}$
is the duration of this transition. The energy density of dark energy
is given by: 
\begin{equation}
\rho_{e}\left(a\right)=\rho_{e}^{0}\exp\left(-3\int_{1}^{a}\frac{\left(1+w\left(a'\right)\right)da'}{a'}\right)\,.
\end{equation}
For all models that we study we assume that the matter density parameter
now is $\Omega_{m}^{0}=0.315$, according to Planck results for a
flat $\Lambda$CDM model \cite{Ade2013k}.

One of the main advantages in the use of the growth index is that
if geometrical probes can tightly constrain the background evolution,
and then $\Omega_{m}\left(a\right)$, different theories of gravitation
can be distinguished via the different values they predict for $\gamma$
\cite{Linder:2005in}. Here we want to show that the impact dark energy
perturbations on $\gamma$ can be as large as in some modified gravity
models. However, as we will soon discuss, since one of the key parameters
that affects dark energy perturbations is $w$, which also affects
the background evolution, in order to make a realistic comparison,
we restrict the evolution of $w$ in a way that Hubble parameter instantaneous
deviation from the $\Lambda$CDM one is no larger than $12\%$.

\begin{figure}
\begin{centering}
\includegraphics[scale=0.9]{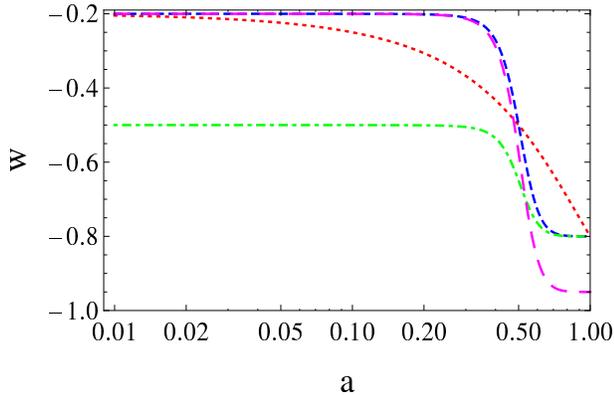}
\par\end{centering}

\caption{Evolution of $w$ as a function of the scale factor, $a$, for model
A (dotted red line), model B (short-dashed blue line), model C (dot-dashed
green line) and model D (long-dashed magenta line). \label{fig:w_hubb}}
\end{figure}

\begin{figure}
\begin{centering}
\includegraphics[scale=0.87]{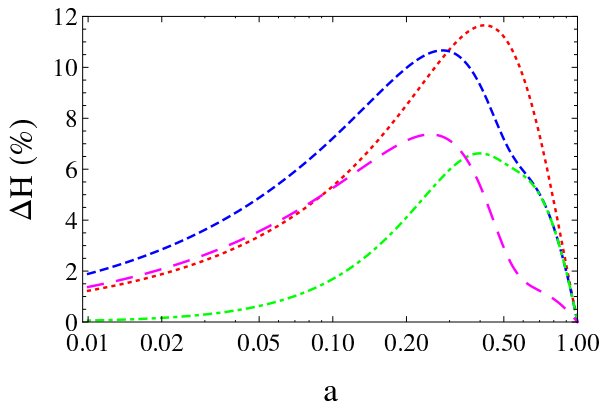}
\par\end{centering}

\caption{Evolution of $\Delta H$, equation \ref{eq:DH}, as a function of
the scale factor, $a$. The line styles are the same as in figure
\ref{fig:w_hubb}. \label{fig:w_hubb-1}}
\end{figure}

Keeping that in mind, we choose four different sets of parameters, which are
shown in table \ref{tab:w_values}. Model A has a smooth transition
from $w_{m}=-0.2$ to $w_{0}=-0.8$, while model B has the same limiting
values, but with a fast transition at low redshift. The transition
in model C is as fast as in model B, but it has a smaller $w_{m}$.
Finally, model D has the same parameters as in model B, but with $w_{0}=-0.95$.
The evolution of $w$ for the four models is shown in figure \ref{fig:w_hubb}.
Assuming the same Hubble constant for all models, in figure \ref{fig:w_hubb-1}
we show the percentage difference of the Hubble parameter of a given
model, $H\left(z\right)_{mod}$, relative to the one in $\Lambda$CDM,
$H\left(z\right)_{\Lambda}$, 
\begin{equation}
\Delta H=100\times\left(\frac{H\left(z\right)_{mod}}{H\left(z\right)_{\Lambda}}-1\right)\%\,.
\label{eq:DH}
\end{equation}

\begin{table}
\begin{ruledtabular}
\begin{tabular}{ccccc}
Model & $w_{0}$ & $w_{m}$ & $a_{c}$ & $\Delta_{m}$\tabularnewline
\hline
A & -0.8 & -0.2 & 0.5 & 4\tabularnewline
B & -0.8 & -0.2 & 0.5 & 0.05\tabularnewline
C & -0.8 & -0.5 & 0.5 & 0.05\tabularnewline
D & -0.95 & -0.2 & 0.5 & 0.05\tabularnewline
\end{tabular}
\end{ruledtabular}
\par

\caption{Table with the parameters for the equation of state of four dark energy
models, described by the parametrization in equation (\ref{eq:corass_parm-1}).\label{tab:w_values}}
\end{table}

At $z=1000$, model C has $\Omega_{e}\sim10^{-5}$, but models A,
B and D have $\Omega_{e}\sim10^{-3}$, therefore these three can be
considered as Early Dark Energy models. We will show that the combination
of large and rapid decay of $w$ at low $z$ can cause important modifications
to the growth of matter perturbations. But let us first define the
system of equations that we will use to evolve the linear perturbations
of matter and dark energy and discuss some relevant features of dark
energy perturbations.

\section{Evolution of perturbations}

As already noted in reference \cite{Dossett:2013npa}, dark energy
models with anisotropic stress are not affected by modifications of
the sound speed. Therefore, here we will focus on dark energy without
anisotropic stress, hence the system of equations that governs the
evolution of matter and dark energy perturbations in the Newtonian
gauge, in Fourier space, is given by \cite{Ma:1995ey}: 

\selectlanguage{brazil}%
\begin{equation}
\dot{\delta}_{m}+\theta_{m}=3\dot{\phi}\;,\label{comp_GR_mat_cont}
\end{equation}
\begin{equation}
\dot{\theta}_{m}+\mathcal{H}\theta_{m}=k^{2}\phi\;,\label{comp_GR_mat_euler}
\end{equation}
\begin{equation}
\dot{\delta}_{e}+3\mathcal{H}\left(\frac{\delta p_{e}}{\delta\rho_{e}}-w\right)\delta_{e}+\left(1+w\right)\theta_{e}=3\left(1+w\right)\dot{\phi}\;,\label{comp_GR_de_cont}
\end{equation}
\begin{equation}
\dot{\theta}_{e}+\mathcal{H}\left(1-3c_{a}^{2}\right)\theta_{e}=\frac{\left(\delta p_{e}/\delta\rho_{e}\right)k^{2}\delta_{e}}{\left(1+w\right)}+k^{2}\phi\;,\label{comp_GR_de_euler}
\end{equation}

\begin{equation}
k^{2}\phi+3\mathcal{H}\left(\dot{\phi}+\mathcal{H}\phi\right)=-\frac{3\mathcal{H}^{2}}{2}\left(\Omega_{m}\delta_{m}+\Omega_{e}\delta_{e}\right)\;,\label{comp_PN_poisson}
\end{equation}
\foreignlanguage{english}{where }
\begin{equation}
c_{a}^{2}=\frac{\dot{p}_{e}}{\dot{\rho}_{e}}=w-\frac{\dot{w}}{3\mathcal{H}\left(1+w\right)}\label{eq:ad_sound}
\end{equation}
\foreignlanguage{english}{is the squared adiabatic sound speed of
dark energy perturbations and the total pressure perturbation is given
by \cite{Bean2004}:
\begin{equation}
\delta p_{e}=c_{{\rm eff}}^{2}\delta\rho_{e}+3\mathcal{H}\left(1+w\right)\left(c_{{\rm eff}}^{2}-c_{a}^{2}\right)\rho_{e}\frac{\theta_{e}}{k^{2}}\,.\label{eq:cov_press}
\end{equation}
We integrate this system from $z=1000$ to $z=0$.}

\selectlanguage{english}%
In the case of negligible $c_{{\rm eff}}$, dark energy perturbations
grow at the same pace of matter perturbations during matter dominated
era \cite{Abramo:2008ip,Sapone2009}:

\begin{equation}
\delta_{e}=\frac{\left(1+w\right)}{\left(1-3w\right)}\delta_{m}\,.\label{de_dm_sol}
\end{equation}
In Early Dark Energy models, which have $w\simeq0$ during matter
dominated era, dark energy perturbations have the same order of magnitude
of matter perturbations. Therefore, in order to set the initial conditions,
we have to take this fact under consideration \cite{Batista:2013oca}.

Solution (\ref{de_dm_sol}) is valid on small scales, during matter
dominated era and for a constant $w$, but is also a good order of
magnitude estimation for $\delta_{e}$, even during the period of
transition to dark energy domination and for varying $w$ \cite{Batista:2013oca}.
Therefore, it is clear that the lower the $|w|$, the larger the $\delta_{e}$.
However, we must have $w\simeq-1$ at low redshift so dark
energy can accelerate the cosmic expansion. Therefore, dark energy perturbations
will necessarily decrease during the transition to the accelerated
expansion. If, however, $w$ has a rapid transition from $w\simeq0$
to $w\simeq-1$ at low redshift, $\delta_{e}$ still can strongly
influence $\delta_{m}$ now. Moreover, during the transition from decelerated
to accelerated expansion $\Omega_{e}$ grows, partially compensating
the decrease of $\delta_{e}$ in equation (\ref{comp_PN_poisson}). 

It is important to highlight that the impact of dark energy perturbations
strongly depends on the evolution of $w$. It has been shown that
two-parameter descriptions of $w$ have strong limitations, particularly
being not able to follow and constrain rapid evolution \cite{Bassett2004a}.
Therefore the use of such parametrizations certainly will not be able to
describe the potential impact of dark energy perturbations on the
matter growth. 

As we will show, the impact of dark energy perturbations on matter
perturbations is strongest in model B, which has a rapid and large
decay of $w$ at low $z$. The impact is weaker in model A because
of its slower transition of $w$, which makes the decay of $\delta_{e}$
occur earlier. For model C the impact is even weaker because it
has a lower value of $w$ during matter dominated era. In model D,
which has the same parameters of model B, but with a value of $w$
much closer to $-1$ now, the impact is only slightly weaker than
in model B. 

If $c_{{\rm eff}}$ is not negligible, dark energy perturbations are
much smaller than matter perturbations on small scales, therefore
their impact on $\delta_{m}$ will be much weaker. This situation
may be considered as smooth or non-clustering dark energy and it is a
feature of nearly all quintessence models, with exception of models
with field potentials that can induce a rapid field oscillation \cite{Amin2012a}.
Indeed, as we will show, when we set $c_{{\rm eff}}=1$ for the models
that we consider, the growth index is very similar to the $\Lambda$CDM
one. For more detailed studies about the phenomenology of dark energy
perturbations, in both clustering and non-clustering situations, see,
for instance, \cite{Abramo:2008ip,Sapone2009,Ballesteros2010b,Sefusatti2011,Batista:2013oca}.

In the following section, we will study the growth index for the two
limiting cases of smooth, $c_{{\rm eff}}=1$, and clustering, $c_{{\rm eff}}=0$,
dark energy models. For all the results of the growth index shown
in the next section, we have fixed the wavenumber to $k=0.2\mbox{ h Mpc}^{-1}$.

\section{Growth index}

\begin{figure}
\begin{centering}
\includegraphics[scale=0.9]{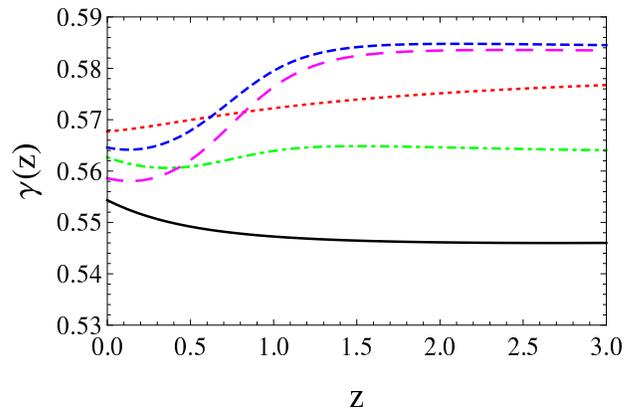}
\par\end{centering}

\caption{Evolution of $\gamma\left(z\right)$, equation (\ref{gma_z}), for
smooth dark energy models: model A (dotted red line), model B (short-dashed
blue line), model C (dot-dashed green line), model D (long-dashed
magenta line) and also for $\Lambda$CDM model (solid black line).\label{fig:gma_z}}
\end{figure}

Once we solve the system of equations (\ref{comp_GR_mat_cont}) -
(\ref{comp_PN_poisson}), we compute $f$ and then determine the growth
index as a function of the redshift: 
\begin{equation}
\gamma\left(z\right)=\frac{\ln f\left(z\right)}{\ln\Omega_{m}\left(z\right)}\,.\label{gma_z}
\end{equation}
In figure \ref{fig:gma_z} we show, for $c_{{\rm eff}}=1$, the evolution
of $\gamma\left(z\right)$ for the models A, B, C and D and also the
results for $\Lambda$CDM model. As is well-known, we see that $\Lambda$CDM
model has a nearly constant value of the growth index $\gamma\simeq0.55$
and the values of $\gamma\left(z\right)$ in models with smooth dark
energy are not too far from it, presenting a mild variation with $z$. 

\begin{figure}
\begin{centering}
\includegraphics[scale=0.9]{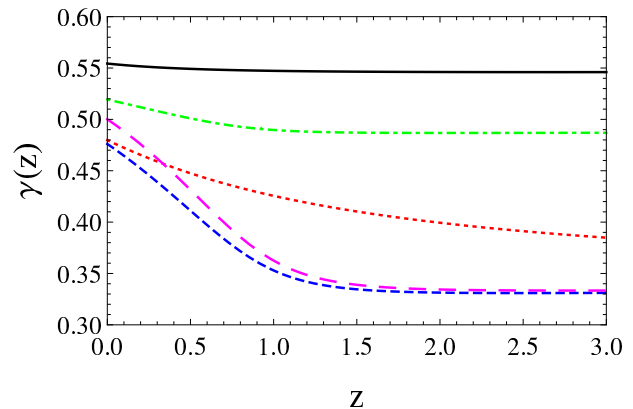}
\par\end{centering}

\caption{Evolution of $\gamma\left(z\right)$, equation (\ref{gma_z}), for
clustering dark energy models: model A (dotted red line), model B
(short-dashed blue line), model C (dot-dashed green line), model D
(long-dashed magenta line) and also for $\Lambda$CDM model (solid
black line).  Note that the range of variation of $\gamma$ is larger than in 
figure \ref{fig:gma_z}. 
\label{fig:gma_z-1}}
\end{figure}

The results for $c_{{\rm eff}}=0$ are shown in figure \ref{fig:gma_z-1}.
We can clearly see that in the case of clustering dark energy, large
deviations from $\Lambda$CDM and smooth dark energy models are present,
including a strong evolution with time, which clearly suggests that
the usual constant growth index parametrization may not be appropriate
in these cases. Since for a given model with smooth or clustering
dark energy the background evolution is the same, it is clear that
dark energy clustering lowers $\gamma$ and induces an important time
variation on it. 

\begin{figure}
\begin{centering}
\includegraphics[scale=0.9]{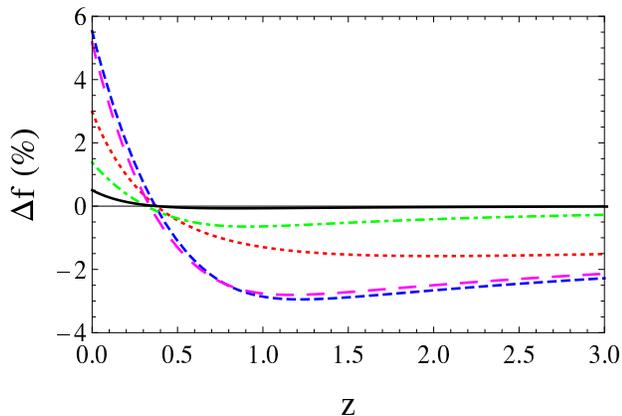}
\par\end{centering}

\caption{Evolution of $\Delta f$, equation (\ref{f_diff}), for the constant
growth index parametrization: $\Lambda$CDM model (solid black line),
model A (dotted red line), model B (short-dashed blue line), model
C (dot-dashed green line) and and Model D (long-dashed magenta line).
\label{fig:f_diff}}
\end{figure}

When we fit the numerical solution of $f$ with a constant growth
index, $\gamma_{0}$, we find that smooth dark energy models have
$\gamma_{0}\simeq0.56$ with an accuracy better than $0.5\%$. For
the clustering models, however, since $\gamma\left(z\right)$ has
a large variation with $z$, the constant $\gamma_{0}$ parametrization
is much more inaccurate. In figure \ref{fig:f_diff} we show, only
for the clustering models, the percentage difference relative to the
numerical solution of $f$,
\begin{equation}
\Delta f=100 \times \left(\frac{f_{p}}{f}-1\right)\%\,,
\label{f_diff}
\end{equation}
where $f_{p}$ is the corresponding parametrized expression. We observe
that the constant $\gamma_{0}$ parametrizations for models A and
C present errors about $2\%$, but reaching almost $6\%$ for
models B and D at very low $z$.

\begin{figure}
\begin{centering}
\includegraphics[scale=0.9]{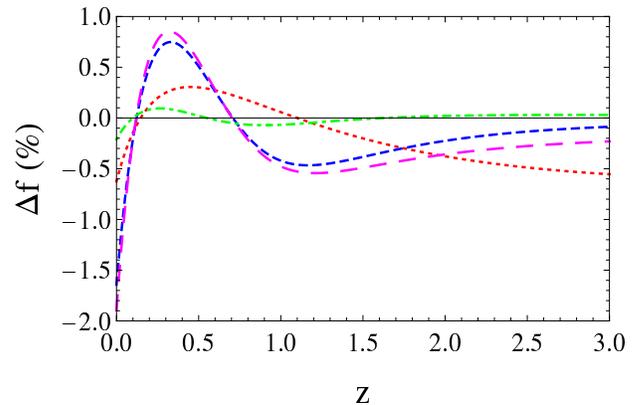}
\par
\end{centering}
\caption{Evolution of $\Delta f$, equation (\ref{f_diff}), for the redshift
dependent $\gamma$ parametrization, equation (\ref{gma_z_par}),
for model A (dotted red line), model B (short-dashed blue line), model
C (dot-dashed green line) and model D (long-dashed magenta line).
With this parametrization $\Delta f$ for $\Lambda$CDM model is bellow
$0.01\%$ level and is indistinguishable from the horizontal axis.
\label{fig:f_diff-1}}
\end{figure}

Given this inaccuracy of the constant $\gamma_{0}$ parametrization
in the clustering models, we also study the following redshift dependent
parametrization for $\gamma$ \cite{Dossett2010}:
\begin{equation}
\gamma\left(z\right)=\gamma_{e}+\gamma_{b}e^{-z/0.61}\,.\label{gma_z_par}
\end{equation}
In figure \ref{fig:f_diff-1} we show that the errors associated with this
parametrization  is much smaller, always bellow the percent level, expect at very low
$z$ in models B and D. Again we note that clustering dark energy
models may not be very accurately described by parametrizations for
the growth index, especially when presenting a large and fast decay
of its equation of state, like in model B and D. 

In table \ref{tab:gma_fits} we present the best fit values for the
constant $\gamma_{0}$ and variable $\gamma$, equation \eqref{gma_z_par},
parametrizations. It is clear that in the presence of clustering dark
energy the values of $\gamma_{0}$ can be quite bellow
the $\Lambda$CDM one. When considering the variable $\gamma$ parametrization,
we can see that the values of $\gamma_{b}$, which is associated with
the time dependence of $\gamma$, are one order of magnitude larger
than in smooth dark energy models, or clustering dark energy with
slowly varying equation of state \cite{Dossett:2013npa}. This significant
time variation of $\gamma$ is actually an important feature of these
clustering dark energy models.

One interesting aspect that can be analyzed with this redshift dependent
parametrization is the sign and magnitude of $\gamma_{b}$, which
provides information about the slope of the growth index. It has been
noted that in General Relativity the slope is positive \cite{Wu2009a,Dossett2010,Dossett:2013npa}.
For all clustering dark energy models that we have considered, the 
slope for the growth index is positive, $\gamma_{b}>0$ (see table \ref{tab:gma_fits}).
Therefore this feature can still be used to discriminate between General
Relativity and DGP models, which have negative slope. Moreover, DGP
model has a large growth index ($\simeq0.68$), whereas in clustering dark energy models
it is much lower, e.g., $\gamma\left(z\right)<0.48$
for models A and B. Hence, if future observations point to high values
of $\gamma$, the indication of modified gravity is clear regardless
of the presence of clustering dark energy. 

On the other hand, if future observations point to low values of $\gamma$,
the indication of modified gravity is not clear anymore. In some $f\left(R\right)$
models, we may have $\gamma\left(z\right)\lesssim0.42$ and with a strong time variation \cite{Gannouji2009,Tsujikawa2009}, a similar evolution to what we have found for clustering
dark energy models. Furthermore, the slope of the growth index in these
$f\left(R\right)$ models is positive, just as we have found for clustering
dark energy. However, since $\gamma$ is basically monotonically decreasing
with $z$ in $f\left(R\right)$, whereas in clustering dark energy
it eventually reaches a minimum (no lower than $0.33$ for models
B and D), precise measurements for intermediate redshifts can in principle
discriminate between these two scenarios. 

Therefore, we have shown that when dark energy is allowed to cluster,
the growth index alone may not be a valid way to distinguish between
General Relativity and modified gravity. However, this result strongly depends
on the time variation of $w$. Then we have to stress that to make dark energy
perturbations cause a strong impact on the growth
rate, dark energy equation of state must have a rapid and large decay
at low $z$. For instance, in model B, the one with largest difference
relative to $\Lambda$CDM, we have a rapid decay from $-0.2$ to $-0.8$
happening around $z=1$. 

\begin{table}
\begin{ruledtabular}
\begin{tabular}{cccc}
Model & $\gamma_{0}$ & $\gamma_{e}$ & $\gamma_{b}$\tabularnewline
\hline
A, $c_{{\rm eff}}=0$ & 0.4544 & 0.4096 & 0.0756\tabularnewline
B, $c_{{\rm eff}}=0$ & 0.4295 & 0.3336 & 0.1568\tabularnewline
C, $c_{{\rm eff}}=0$ & 0.5074 & 0.4844 & 0.0367\tabularnewline
D, $c_{{\rm eff}}=0$ & 0.4564 & 0.3453 & 0.1714\tabularnewline
A, $c_{{\rm eff}}=1$ & 0.5697 & 0.5737 & -0.0069\tabularnewline
B, $c_{{\rm eff}}=1$ & 0.5684 & 0.5807 & -0.0206\tabularnewline
C, $c_{{\rm eff}}=1$ & 0.5618 & 0.5628 & -0.0017\tabularnewline
D, $c_{{\rm eff}}=1$ & 0.5619 & 0.5764 & -0.0227\tabularnewline
\end{tabular}
\end{ruledtabular}
\par 

\caption{Table with the fitted parameters for $f$ with the constant growth
index parametrization, $\gamma_{0}$, and the redshift dependent parametrization,
equation (\ref{gma_z}) .\label{tab:gma_fits}}
\end{table}

Another interesting point that can be analyzed with the growth index
is its potential ability to distinguish between smooth and clustering
dark energy models. In the context of Early Dark Energy models it
has been shown that galaxy cluster abundances in clustering dark energy
models are more similar to the $\Lambda$CDM one than the non-clustering
models with the same background evolution \cite{Batista:2013oca}.
As we can see in figure \ref{fig:gma_z-1}, dark energy perturbations
strongly impact the values and redshift dependence of the growth index,
showing a clear distinction from the smooth dark energy models, figure
\ref{fig:gma_z}.

\section{Conclusions}

We have shown that clustering dark energy models with a rapid and
large decay of equation of state at low redshift have a growth index
much lower than in $\Lambda$CDM or quintessence models. When fitting
a constant growth index, we found values in the range $0.43<\gamma_{0}<0.51$
for the four representative clustering models that we have considered.
However, we also have observed that the constant $\gamma_{0}$ fit
is not very accurate to represent the numerical solution for the growth
rate, $f$, being almost $6\%$ inaccurate for our models B and D. 

When fitting $f$ with the redshift dependent parametrization for $\gamma$ 
proposed in reference \cite{Dossett2010}, the accuracy is about four
times better, but can still be as large as $1\%$. In this parametrization,
the $\gamma_{b}$ parameter, associated with the time dependence of
$\gamma$, is always positive, and one order of magnitude larger
for clustering dark energy models when compared to their non-clustering
versions. This behavior can be useful in order to distinguish between
smooth and clustering dark energy.

The general trend we have found is that clustering dark energy lowers
$\gamma$ and induces an important time variation to it. This is the
same behavior found in some $f\left(R\right)$ models \cite{Gannouji2009,Tsujikawa2009}.
Therefore, we conclude that if future data on the growth index points
to values much lower than in the $\Lambda$CDM model, the interpretation
of the result as a clear evidence of modified gravity is not straightforward.

\subsection*{Acknowledgments}
RCB thanks Jailson Alcaniz for useful discussions and FAPERN for the
financial support.

\bibliographystyle{apsrev}
\bibliography{referencias}

\end{document}